\documentclass[11pt,a4paper,twoside]{article}
\usepackage[sc]{mathpazo}
\usepackage[english]{babel}
\usepackage{a4wide}
\usepackage{amsmath,amssymb}
\usepackage{epsfig}
\usepackage{graphicx}
\usepackage{subfigure}
\usepackage{xspace}
\usepackage[margin=\parindent,bf]{caption}
\usepackage{units}
\usepackage{booktabs}
\usepackage{multirow}
\usepackage{multicol}
\usepackage{shadow}
\usepackage[pdftex,dvipsnames,hyperref]{xcolor}
\usepackage[pdftex,colorlinks=true,linkcolor=BrickRed,citecolor=ForestGreen,urlcolor=Blue]{hyperref}

\def\gap#1#2{GAP-{#1}-{#2}}
\def\HRule{\rule{\linewidth}{1mm}}
\def\eq#1{\begin{equation}#1\end{equation}}
\def\al#1{\begin{align}#1\end{align}}

\def\fig#1#2#3{\begin{figure}[t]\centering{#1}\caption[]{#2}%
\label{#3}\end{figure}}
\def\W{\operatorname{W}}
\def\figw{0.6\textwidth}
\def\figwt{0.48\textwidth}

\def\O#1{\mathcal{O}(#1)}

\def\eps{\varepsilon}
\def\ie{e^{-1}}

\begin{document}

\pagestyle{empty}
\vspace*{\stretch{1}}
\begin{flushright}
  \HRule
  \\[9mm]
  \large
  $~$
  \\[7mm]
  {\bf\huge Having Fun with Lambert $\boldsymbol{\W(x)}$ Function}
  \\[10mm]
  \hfill%
  \parbox[b]{10cm}{\begin{flushright}
    Darko Veberi\v{c} $^\text{a,b,c}$
    \\[6mm]
    $^\text{a}$ University of Nova Gorica, Slovenia\\
    $^\text{b}$ IK, Forschungszentrum Karlsruhe, Germany\\
    $^\text{c}$ J. Stefan Institute, Ljubljana, Slovenia
    \\[10mm]
    June 2009
  \end{flushright}}
  \\[5mm]
  \HRule
\end{flushright}
\vspace*{\stretch{1}}
\begin{abstract}
This short note presents the Lambert $\W(x)$ function and its possible 
application in the framework of physics related to the Pierre Auger 
Observatory. The actual numerical implementation in C++ consists of 
Halley's and Fritsch's iteration with branch-point expansion, asymptotic 
series and rational fits as initial approximations.
\end{abstract}
\vspace*{\stretch{3}}
\clearpage

\section{Introduction}

\fig{\includegraphics[width=\figw]{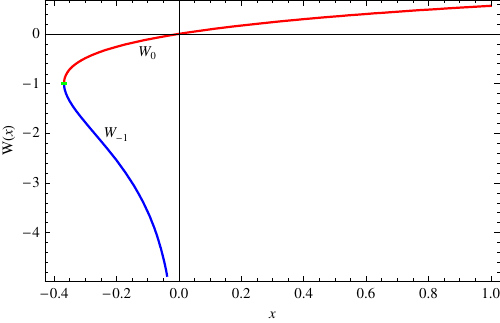}}
{The two branches of the Lambert W function, $\W_{-1}(x)$ in blue and 
$W_0(x)$ in red. The branching point at $(-\ie,\,-1)$ is denoted with 
a green dash.}
{f:lambertW}

The Lambert $\W(x)$ function is defined as the inverse function of
\eq{
y\exp y = x,
\label{def1}
}
the solution being given by
\eq{
y=\W(x),
\label{def2}
}
or shortly
\eq{
\W(x)\exp\W(x)=x.
\label{definition}
}

Since the $x\mapsto x\exp x$ mapping is not injective, no unique inverse 
of the $x\exp x$ function exists. As can be seen in 
Fig.~\ref{f:lambertW}, the Lambert function has two real branches with a 
branching point located at $(-\ie,\,-1)$. The bottom branch, 
$\W_{-1}(x)$, is defined in the interval $x\in[-\ie,\,0]$ and has a 
negative singularity for $x\to0^-$. The upper branch is defined for 
$x\in[-\ie,\,\infty]$.

The earliest mention of problem of Eq.~\eqref{def1} is
attributed to Euler. However, Euler himself credited Lambert for his 
previous work in this subject. The $\W(x)$ function started to be named 
after Lambert only recently, in the last 10 years or so. The letter $\W$ 
was chosen by the first implementation of the $\W(x)$ function in the 
Maple computer software.

Recently, the $\W(x)$ function amassed quite a following in the
mathematical community. Its most faithful proponents are suggesting to
elevate it among the present set of elementary functions, such as
$\sin(x)$, $\cos(x)$, $\ln(x)$, etc. The main argument for doing so is 
the fact that it is the root of the simplest exponential polynomial 
function.

While the Lambert W function is simply called W in the mathematics software
tool \emph{Maple}, in the \emph{Mathematica} computer algebra framework
this function is implemented under the name \texttt{ProductLog} (in the 
recent versions an alias \texttt{LambertW} is also supported).

There are numerous, well documented applications of $\W(x)$ in mathematics,
physics, and computer science \cite{knuth,corless}. Here we will give two
examples that arise from the physics related to the Pierre Auger 
Observatory.

\subsection{Moyal function}

\fig{
\includegraphics[width=\figwt]{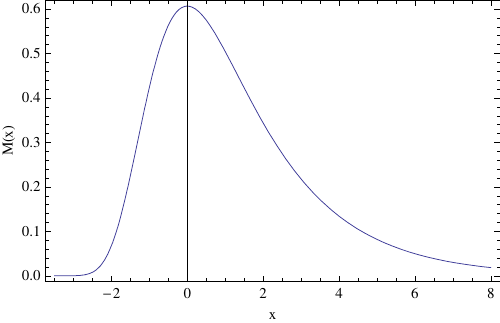}
\quad
\includegraphics[width=\figwt]{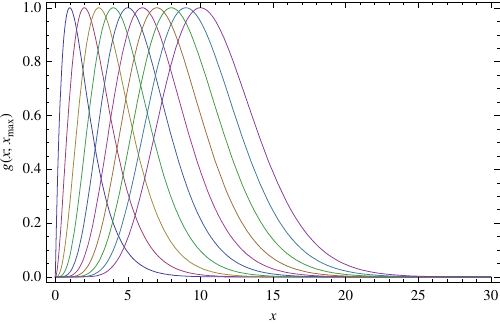}
}{\emph{Left:} The Moyal function $\operatorname{M}(x)$. \emph{Right:} A 
family of one-parametric Gaisser-Hillas functions $g(x;\,x_\text{max})$
for $x_\text{max}$ in the range from 1 to 10 with step 1.}
{f:moyal-gh}

Moyal function is defined as
\eq{
\operatorname{M}(x)=\exp\left(-\tfrac12\left(x+\exp(-x)\right)\right).
}
Its inverse can be written in terms of the two branches of the Lambert W
function,
\eq{
\operatorname{M}^{-1}_\pm(x)=
  \W_{0,-1}(-x^2) - 2\ln x.
}
and can be seen in Fig.~\ref{f:moyal-gh} (left).

Within the event reconstruction of the data taken by the Pierre Auger 
Observatory, the Moyal function is used for phenomenological recovery of 
the saturated signals from the photomultipliers.

\subsection{Gaisser-Hillas function}

In astrophysics the Gaisser-Hillas function is used to model the 
longitudinal particle density in a cosmic-ray air showers \cite{gh}. We 
can show that the inverse of the three-parametric Gaisser-Hillas 
function,
\eq{
G(X;\,X_0,X_\text{max},\lambda) =
  \left[
    \frac{X-X_0}{X_\text{max}-X_0}
  \right]^{\frac{X_\text{max}-X_0}\lambda}
  \exp\left(\frac{X_\text{max}-X}\lambda\right),
}
is intimately related to the Lambert W function. Using rescale substitutions,
\al{
x &= \frac{X-X_0}\lambda\qquad\text{and}
\\
x_\text{max} &= \frac{X_\text{max}-X_0}\lambda,
}
the Gaisser-Hillas function is modified into a function of one parameter 
only,
\eq{
g(x;\,x_\text{max}) =
  \left[\frac{x}{x_\text{max}}\right]^{x_\text{max}}
  \exp(x_\text{max}-x).
}
The family of one-parametric Gaisser-Hillas functions is shown in 
Fig.~\ref{f:moyal-gh} (right). The problem of finding an inverse,
\eq{
g(x;\,x_\text{max}) \equiv a
}
for $0 < a \leqslant 1$, can be rewritten into
\eq{
-\frac{x}{x_\text{max}}\exp\left(-\frac{x}{x_\text{max}}\right) =
  -a^{1/x_\text{max}}\,\ie.
}
According to the definition \eqref{def1}, the two (real) solutions for 
$x$ are obtained from the two branches of the Lambert W function,
\eq{
x_{1,2} =
  -x_\text{max}\W_{0,-1}(-a^{1/x_\text{max}}\,\ie) =
  -x_\text{max}\W_{0,-1}(-\sqrt[x_\text{max}]{a}/e).
}
Note that the branch $-1$ or $0$ simply chooses the right or left side 
relative to the maximum, respectively.

\section[]{Numerics}

Before moving to the actual implementation let us review some of the 
possible nimerical and analytical approaches.

\subsection{Recursion}

For $x>0$ and $\W(x)>0$ we can take the natural logarithm of 
\eqref{definition} and rearrange it,
\eq{
\W(x) = \ln x  - \ln\W(x).
\label{rear}
}
It is clear, that a possible analytical expression for $\W(x)$
exhibits a degree of self similarity. The $\W(x)$ function has multiple
branches in the complex domain. Due to the $x>0$ and $\W(x)>0$ 
conditions, the Eq.~\eqref{rear} represents the positive part of the 
$\W_0(x)$ principal branch, but as it turns out, in this form it is 
suitable for evaluation when $\W_0(x)>1$, i.e.\ when $x>e$.

Unrolling the self-similarity \eqref{rear} as a recursive relation, one 
obtains the following curious expression for $\W_0(x)$,
\eq{
\W_0(x) = \ln x - \ln(\ln x - \ln(\ln x - \,\ldots\,)),
\label{cle0}
}
or in a shorthand of a continued logarithm,
\eq{
\W_0(x) = \ln\frac{x}{\ln\frac{x}{\ln\frac{x}{\cdots}}}.
}
The above expression is clearly a form of successive approximation, the
final result given by the limit, when it exists.

For $x<0$ and $\W(x)<0$ we can multiply both sides of 
Eq.~\eqref{definition} with $-1$, take logarithm, and rewrite it to get 
a similar expansion for the $\W_{-1}(x)$ branch,
\eq{
\W(x) = \ln(-x) - \ln(-\W(x)).
}
Again, this leads to a similar recursive expression,
\eq{
\W_{-1}(x) = \ln(-x) - \ln(-(\ln(-x) - \ln(-(\ln(-x) - \,\ldots\,)))),
\label{clem1}
}
or as a continued logarithm,
\eq{
\W_{-1}(x) = \ln\frac{-x}{-\ln\frac{-x}{-\ln\frac{-x}{\cdots}}}.
\label{rec-w-1}
}
For this continued logarithm we will use the symbol $R_{-1}^{[n]}(x)$ 
where $n$ denotes the depth of the recursion.

Starting from yet another rearrangement of Eq.~\eqref{definition},
\eq{
\W(x)=\frac{x}{\exp\W(x)},
}
we can obtain a recursion relation for the $-\ie<x<e$ part of the 
principal branch $\W_0(x)<1$,
\eq{
\W_0(x) = \frac{x}{\exp\frac{x}{\exp\frac{x}{\ldots}}}.
}

\subsection{Halley's iteration}

We can apply Halley's root-finding method \cite{scavo} to derive an 
iteration scheme for $f(y)=W(y)-x$ by writing the second-order Taylor 
series
\eq{
f(y) = f(y_n) + f'(y_n)\,(y - y_n) + \tfrac12 f''(y_n)\,(y - y_n)^2
       + \cdots
\label{halley-second}
}
Since root $y$ of $f(y)$ satisfies $f(y)=0$ we can approximate the 
left-hand side of Eq.~\eqref{halley-second} with 0 and replace $y$ with 
$y_{n+1}$.  Rewriting the obtained result into
\eq{
y_{n+1} = y_n - \frac{f(y_n)}{f'(y_n)
          + \tfrac12 f''(y_n)\,(y_{n+1}-y_n)}
}
and using Newton's method $y_{n+1}-y_n=-f(y_n)/f''(y_n)$ on the last 
bracket, we arrive at the expression for the Halley's iteration for 
Lambert function
\eq{
W_{n+1} = W_n + \frac{t_n}{t_n\,s_n - u_n},
}
where
\al{
t_n &= W_n\exp W_n - x,
\\
s_n &= \frac{W_n+2}{2(W_n+1)},
\\
u_n &= (W_n+1)\exp W_n.
}

This method is of the third order, i.e.\ having $W_n=\W(x)+\O\eps$ will 
give $W_{n+1}=\W(x)+\O{\eps^3}$. Supplying this iteration with 
sufficiently accurate first approximation of the order of $\O{10^{-4}}$ 
will thus give a machine-size floating point precission $\O{10^{-16}}$ 
in at least two iterations.

\subsection{Fritsch's iteration}

For both branches of Lambert function a more efficient iteration scheme 
exists \cite{fritsch},
\eq{
W_{n+1}=W_n(1+\eps_n),
}
where $\eps_n$ is the relative difference of successive approximations at 
iteration $n$,
\eq{
\eps_n=\frac{W_{n+1}-W_n}{W_n}.
}
The relative difference can be expressed as
\eq{
\eps_n = 
  \left(\frac{z_n}{1+W_n}\right)
  \left(\frac{q_n-z_n}{q_n-2z_n}\right),
}
where
\al{
z_n&=\ln\frac{x}{W_n}-W_n,
\\
q_n&=2(1+W_n)\left(1+W_n+\tfrac23z_n\right).
}
The error term in this iteration is of a fourth order, i.e. with 
$W_n=\W(x)+\O{\eps_n}$ we get $W_{n+1}=\W(x)+\O{\eps_n^4}$.

Supplying this iteration with a sufficiently reasonable first guess,
accurate to the order of $\O{10^{-4}}$, will therefore deliver
machine-size floating point precission $\O{10^{-16}}$ in only one 
iteration and excessive $\O{10^{-64}}$ in two! We have to find reliable 
first order approximation that can be fed into the Fritsch iteration. Due 
to the lively landscape of the Lambert function properties, the 
approximations will have to be found in all the particular ranges of the 
function behavior.

\section{Initial approximations}

The following section deals with finding the appropriate initial
approximations in the whole definition ranges of the two branches of the
Lambert function.

\subsection{Branch-point expansion}

\fig{
\includegraphics[width=\figwt]{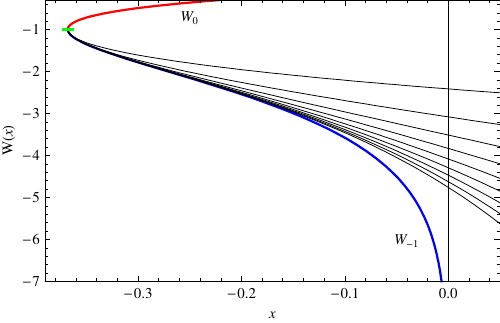}
\quad
\includegraphics[width=\figwt]{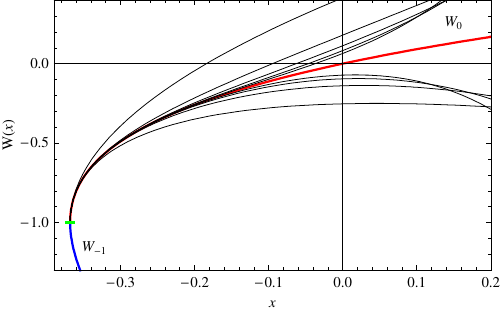}
}{Successive orders of the branch-point expansion for the $\W_{-1}(x)$ on 
the left and $\W_0(x)$ on the right.}
{f:bp-exp}

The inverse of the Lambert function, $\W^{-1}(y)=y\exp y$, has
two extrema located at $\W^{-1}(-1)=-\ie$ and $\W^{-1}(-\infty)=0^-$.  
Expanding $\W^{-1}(y)$ to the second order around the minimum at $y=-1$ 
we obtain
\eq{
\W^{-1}(y) \approx -\frac{1}{e}+\frac{(y+1)^2}{2e}.
}
The inverse $\W^{-1}(y)$ is thus in the lowest order approximated with a 
parabolic term implying that the Lambert function will have square-root 
behavior in the vicinity of the branch point $x=-\ie$,
\eq{
\W_{-1,0}(x) \approx -1 \mp \sqrt{2(1+ex)}.
\label{bp-first}
}
To obtain the additional terms in expression \eqref{bp-first} we proceed 
by defining an inverse function, centered and rescaled around the 
minimum, i.e.\ $f(y)=2(e\W^{-1}(y-1)+1)$. Due to the centering and 
rescaling the Taylor series of this function around $y=0$ becomes 
particularly neat,
\eq{
f(y) = y^2 + \tfrac23y^3 + \tfrac14y^4 + \tfrac1{15}y^5 + \cdots
}
Using this Taylor expansion we can derive coefficients \cite{morse} of 
the corresponding inverse function
\al{
f^{-1}(z) &= 1 + \W\left(\frac{z-2}{2e}\right) =
\label{bp-inverse}
\\
  &= z^{1/2} - \tfrac13z + \tfrac{11}{72}z^{3/2} - \tfrac{43}{540}z^2 + 
     \cdots
}
From Eq.~\eqref{bp-inverse} we see that $z=2(1+ex)$. Using 
$p_\pm(x)=\pm\sqrt{2(1+ex)}$ as independent variable we can write this 
series expansion as
\eq{
\W_{-1,0}(x) \approx
  B_{-1,0}^{[n]}(x) =
    \sum_{i=0}^n b_i p_\mp^i(x),
\label{bp}
}
where the lowest few coefficients $b_i$ are
\begin{center}
\begin{tabular}{lrrrrrrrrrr}
\toprule
$i$ & 0 & 1 & 2 & 3 & 4 & 5 & 6 & 7 & 8 & 9
\\
\midrule
$b_i$ & -1 & 1 & $-\tfrac13$ & $\tfrac{11}{72}$ & $-\tfrac{43}{540}$ &
        $\tfrac{769}{17\,280}$ & $-\tfrac{221}{8\,505}$ & 
        $\tfrac{680\,863}{43\,545\,600}$ & $-\tfrac{1963}{204\,120}$ &
        $\tfrac{226\,287\,557}{37\,623\,398\,400}$
\\
\bottomrule
\end{tabular}
\end{center}

\subsection{Asymptotic series}

Another useful tool is the asymptotic expansion \cite{debruijn} where 
using
\eq{
A(a,b) = a - b + \sum_k \sum_m C_{km}a^{-k-m-1}b^{m+1}
}
where $C_{km}$ are related to the Stirling number of the first kind, the 
Lambert function can be expressed as
\eq{
\W_{-1,0}(x) = A(\ln(\mp x),\ln(\mp\ln(\mp x)))
}
with $a=\ln x$, $b=\ln\ln x$ for the $\W_0$ branch and $a=\ln(-x)$, 
$b=\ln(-\ln(-x))$ for the $\W_{-1}$ branch. The first few terms are
\al{
A(a,b) & =
  a - b + \frac{b}{a} + \frac{b(-2+b)}{2a^2} +
  \frac{b(6-9b+2b^2)}{6a^3} +
\label{asym}
\\
  & +
  \frac{b(-12+36b-22b^2+3b^3)}{12a^4} +
  \frac{b(60-300b+350b^2-125b^3+12b^4)}{60a^5} + \cdots
\nonumber
}

\subsection{Rational fits}

A useful quick-and-dirty approach to the functional approximation is to 
generate large enough sample of data points $\{w_i\exp w_i,\,w_i\}$.
These points evidently lie on the Lambert function. Within some 
appropriately chosen range of $w_i$ values the points are fitted
with a rational approximation
\eq{
Q(x)=\frac{\sum_i a_i x^i}{\sum_i b_i x^i},
}
varying the order of the polynomials in the nominator and denominator, 
and choosing the one that has the lowest maximal absolute residual in a 
particular interval of interest.

For the $\W_0(x)$ branch, the first set of equally-spaced $w_i$ component 
was chosen in a range that produced $w_i\exp w_i$ values in an interval 
$[-0.3,\,0]$. The optimal rational fit turned out to be
\eq{
Q_0(x) =
 x \frac{1+a_1x+a_2x^2+a_3x^3+a_4x^4}
        {1+b_1x+b_2x^2+b_3x^3+b_4x^4}
\label{rat}
}
where the coefficients for this first approximation $Q_0^{[1]}(x)$ are
\begin{center}
\begin{tabular}{lrrrr}
\toprule
$i$ & 1 & 2 & 3 & 4
\\
\midrule
$a_i$ &
5.931375839364438 & 11.39220550532913 & \phantom{1}7.33888339911111 &
0.653449016991959
\\
$b_i$ &
6.931373689597704 & 16.82349461388016 & 16.43072324143226 &
5.115235195211697
\\
\bottomrule
\end{tabular}
\end{center}

For the second fit of the $\W_0(x)$ branch a $w_i$ range was chosen so 
that the $w_i\exp w_i$ values were produced in the interval $[0.3,\,2e]$ 
giving rise to the second optimal rational fit $Q_0^{[2]}(x)$ of the 
same form as in Eq.~\eqref{rat} but with coefficients
\begin{center}
\begin{tabular}{lrrrr}
\toprule
$i$ & 1 & 2 & 3 & 4
\\
\midrule
$a_i$ &
2.445053070726557 & 1.343664225958226 & 0.148440055397592 &
0.0008047501729130
\\
$b_i$ &
3.444708986486002 & 3.292489857371952 & 0.916460018803122 &
0.0530686404483322
\\
\bottomrule
\end{tabular}
\end{center}

For the $\W_{-1}(x)$ branch one rational approximation of the form
\eq{
Q_{-1}(x) =
  \frac{a_0+a_1x+a_2x^2}
       {1+b_1x+b_2x^2+b_3x^3+b_4x^4+b_5x^5}
\label{rat-1}
}
with the coefficients
\begin{center}
\begin{tabular}{lrrrr}
\toprule
$i$ & 0 & 1 & 2 & 3
\\
\midrule
$a_i$ &
-7.81417672390744 & 253.88810188892484 & 657.9493176902304
\\
$b_i$ &
 & -60.43958713690808 & 99.9856708310761 & 682.6073999909428
\\
\midrule
$i$ & 4 & 5
\\
\midrule
$b_i$ & 962.1784396969866 & 1477.9341280760887
\\
\bottomrule
\end{tabular}
\end{center}
is enough.

\section{Implementation}

\textcolor{red}{Note that the most recent and up-to-date implementation
\cite{cpc} can be found at:}
\\
\href{https://github.com/DarkoVeberic/LambertW}{\tt https://github.com/DarkoVeberic/LambertW}
\\

To quantify the accuracy of a particular approximation $\widetilde\W(x)$ 
of the Lambert function $\W(x)$ we can introduce a quantity $\Delta(x)$ 
defined as
\eq{
\Delta(x)=-\log_{10}|\widetilde\W(x)-\W(x)|,
}
so that it gives us a number of correct decimal places the approximation 
$\widetilde\W(x)$ is producing for some parameter $x$.

In Fig.~\ref{f:approx-w0} all mentioned approximations for the $\W_0(x)$ 
are shown in the linear interval $[-\ie,\,0.3]$ on the left and 
logarithmic interval $[0.3,\,10^5]$ on the right. For each of the 
approximations an use interval is chosen so that the number of accurate 
decimal places is maximized over the whole definition range. For the 
$\W_0(x)$ branch the resulting piecewise approximation
\eq{
\widetilde\W_0(x)=
\begin{cases}
B_0^{[9]}(x) & ;\,-\ie \leqslant x < -0.32358170806015724
\\
Q_0^{[1]}(x) & ;\,-0.32358170806015724 \leqslant x < 0.14546954290661823
\\
Q_0^{[2]}(x) & ;\,0.14546954290661823 \leqslant x < 8.706658967856612
\\
A_0(x) & ;\,8.706658967856612 \leqslant x < \infty
\end{cases}
}
is accurate in the definition range $[-\ie,\,7]$ to at least 5 decimal 
places and to at least 3 decimal places in the whole definition range.  
The $B_0^{[9]}(x)$ is from Eq.~\eqref{bp}, $Q_0^{[1]}(x)$ and 
$Q_0^{[2]}(x)$ are from Eq.~\eqref{rat}, and $A_0(x)$ is from 
Eq.~\eqref{asym}.

\fig{
\includegraphics[width=\figwt]{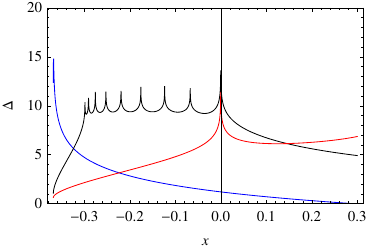}
\quad
\includegraphics[width=\figwt]{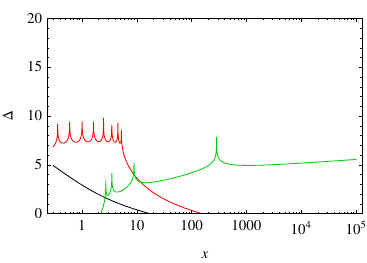}
}{Combining different approximations of $\W_0(x)$ into final piecewise
function. The number of accurate decimal places $\Delta(x)$ is shown for
two ranges, linear interval $[-\ie,\,0.3]$ on the left and
logarithmic interval $[0.3,\,10^5]$ on the right. The
approximation are branch-point expansion $B_0^{[9]}(x)$ from
Eq.~\eqref{bp} in blue, rational fits $Q_0^{[1]}(x)$ and $Q_0^{[2]}(x)$
from Eq.~\eqref{rat} in black and red, respectively, and asymptotic
series $A_0(x)$ from Eq.~\eqref{asym} in green.}
{f:approx-w0}

\fig{
\includegraphics[width=\figwt]{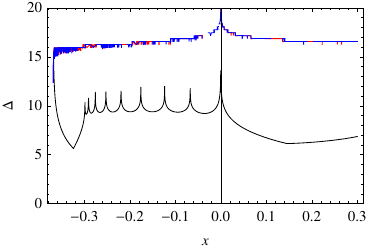}
\quad
\includegraphics[width=\figwt]{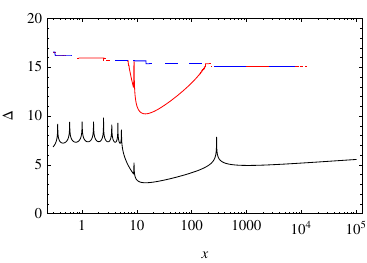}
}{Final values of the combined approximation $\widetilde\W_0(x)$ (black) 
from Fig.~\ref{f:approx-w0} after one step of the Halley iteration (red) 
and one step of the Fritsch iteration (blue).}
{f:approx-w0-step}

The final piecewise approximation $\widetilde\W_0(x)$ is shown in 
Fig.~\ref{f:approx-w0-step} in black line. Using this approximation a 
single step of the Halley iteration (in red) and the Fritsch iteration 
(in blue) is performed and the resulting number of accurate decimal 
places is shown. As can be seen both iterations produce machine-size 
accurate floating point numbers in the whole definition interval except 
for the $[9,110]$ interval where the Halley method requires another step 
of the iteration. For that reason we have decided to use only (one step 
of) the Fritsch iteration in the C++ implementation of the Lambert 
function.

\fig{
\includegraphics[width=\figwt]{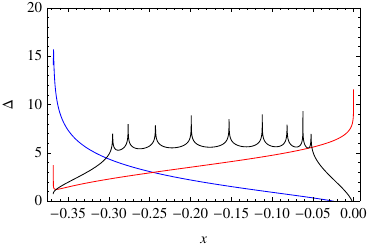}
\quad
\includegraphics[width=\figwt]{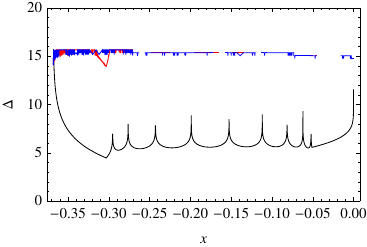}
}{\emph{Left:} Approximations of the $\W_{-1}(x)$ branch. The branch 
point expansion $B_{-1}^{[9]}(x)$ is shown in blue, the rational 
approximation $Q_{-1}(x)$ in black, and the logarithmic recursion 
$R_{-1}^{[9]}$ in red. \emph{Right:} Combined approximation is accurate 
to at least 5 decimal places in the whole definition range. The results 
after applying one step of the Halley iteration are shown in red and 
after one step of the Fritsch iteration in blue.}
{f:approx-w-1}

In Fig.~\ref{f:approx-w-1} (left) the same procedure is shown for the 
$\W_{-1}(x)$ branch. The final approximation
\eq{
\widetilde\W_{-1}(x)=
\begin{cases}
B_{-1}^{[9]}(x) & ;\, -\ie \leqslant x < -0.30298541769
\\
Q_{-1}(x) & ;\, -0.30298541769 \leqslant x < -0.051012917658221676
\\
R_{-1}^{[9]}(x) & ;\, -0.051012917658221676 \leqslant x < 0
\end{cases}
}
is accurate to at least 5 decimal places in the whole definition range 
$[-\ie,\,0]$ and where $B_{-1}^{[9]}(x)$ is from Eq.~\eqref{bp}, 
$Q_{-1}(x)$ is from Eq.~\eqref{rat-1}, and $R_{-1}^{[9]}(x)$ is from 
Eq.~\eqref{rec-w-1}.

In Fig.~\ref{f:approx-w-1} (right) the combined approximation 
$\widetilde\W_{-1}(x)$ is shown in black line. The values after one step 
of the Halley iteration are shown in red and after one step of the 
Fritsch iteration in blue. Similarly as for the previous branch, the 
Fritsch iteration is superior, yielding machine-size accurate results in 
the whole definition range, while the Halley is accurate to at least 13 
decimal places.

\clearpage

\appendix

\section{Implementation in C++}

\textcolor{red}{Note that this is the 2010 version of the implementation. The most up-to-date sources can be found at}
\href{https://github.com/DarkoVeberic/LambertW}{\tt https://github.com/DarkoVeberic/LambertW}

\begin{multicols}{2}

\subsection{\tt Lambert.h}

{\tiny
\begin{verbatim}
/*
  Implementation of Lambert W function

  Copyright (C) 2009 Darko Veberic, darko.veberic@kit.edu

  This program is free software: you can redistribute it and/or modify
  it under the terms of the GNU General Public License as published by
  the Free Software Foundation, either version 3 of the License, or
  (at your option) any later version.

  This program is distributed in the hope that it will be useful,
  but WITHOUT ANY WARRANTY; without even the implied warranty of
  MERCHANTABILITY or FITNESS FOR A PARTICULAR PURPOSE.  See the
  GNU General Public License for more details.

  You should have received a copy of the GNU General Public License
  along with this program.  If not, see <http://www.gnu.org/licenses/>.

  25 Jun 2009
*/

#ifndef _utl_LambertW_h_
#define _utl_LambertW_h_


/** Approximate Lambert W function
  Accuracy at least 5 decimal places in all definition range.
  See LambertW() for details.

  \param branch: valid values are 0 and -1
  \param x: real-valued argument \f$\geq-1/e\f$
  \ingroup math
*/

template<int branch>
double LambertWApproximation(const double x);


/** Lambert W function

  Lambert function \f$y={\rm W}(x)\f$ is defined as a solution
  to the \f$x=ye^y\f$ expression and is also known as
  "product logarithm". Since the inverse of \f$ye^y\f$ is not
  single-valued, the Lambert function has two real branches
  \f${\rm W}_0\f$ and \f${\rm W}_{-1}\f$.

  \f${\rm W}_0(x)\f$ has real values in the interval
  \f$[-1/e,\infty]\f$ and \f${\rm W}_{-1}(x)\f$ has real values
  in the interval \f$[-1/e,0]\f$.
  Accuracy is the nominal double type resolution
  (16 decimal places).

  \param branch: valid values are 0 and -1
  \param x: real-valued argument \f$\geq-1/e\f$ (range depends on 
  branch)
*/

template<int branch>
double LambertW(const double x);


#endif
\end{verbatim}
}

\subsection{\tt Lambert.cc}

{\tiny
\begin{verbatim}
/*
  Implementation of Lambert W function

  Copyright (C) 2009 Darko Veberic, darko.veberic@kit.edu

  This program is free software: you can redistribute it and/or modify
  it under the terms of the GNU General Public License as published by
  the Free Software Foundation, either version 3 of the License, or
  (at your option) any later version.

  This program is distributed in the hope that it will be useful,
  but WITHOUT ANY WARRANTY; without even the implied warranty of
  MERCHANTABILITY or FITNESS FOR A PARTICULAR PURPOSE.  See the
  GNU General Public License for more details.

  You should have received a copy of the GNU General Public License
  along with this program.  If not, see <http://www.gnu.org/licenses/>.
*/

#include <iostream>
#include <cmath>
#include <limits>
#include "LambertW.h"

using namespace std;


namespace LambertWDetail {


  const double kInvE = 1/M_E;


  template<int n>
  inline double BranchPointPolynomial(const double p);


  template<>
  inline
  double
  BranchPointPolynomial<1>(const double p)
  {
    return
      -1 + p;
  }


  template<>
  inline
  double
  BranchPointPolynomial<2>(const double p)
  {
    return
      -1 + p*(1 + p*(-1./3));
  }


  template<>
  inline
  double
  BranchPointPolynomial<3>(const double p)
  {
    return
      -1 + p*(1 + p*(-1./3 + p*(11./72)));
  }


  template<>
  inline
  double
  BranchPointPolynomial<4>(const double p)
  {
    return
      -1 + p*(1 + p*(-1./3 + p*(11./72 + p*(-43./540))));
  }


  template<>
  inline
  double
  BranchPointPolynomial<5>(const double p)
  {
    return
      -1 + p*(1 + p*(-1./3 + p*(11./72 + p*(-43./540 + 
      p*(769./17280)))));
  }


  template<>
  inline
  double
  BranchPointPolynomial<6>(const double p)
  {
    return
      -1 + p*(1 + p*(-1./3 + p*(11./72 + p*(-43./540 + p*(769./17280
      + p*(-221./8505))))));
  }


  template<>
  inline
  double
  BranchPointPolynomial<7>(const double p)
  {
    return
      -1 + p*(1 + p*(-1./3 + p*(11./72 + p*(-43./540 + p*(769./17280
      + p*(-221./8505 + p*(680863./43545600)))))));
  }


  template<>
  inline
  double
  BranchPointPolynomial<8>(const double p)
  {
    return
      -1 + p*(1 + p*(-1./3 + p*(11./72 + p*(-43./540 + p*(769./17280
      + p*(-221./8505 + p*(680863./43545600 + p*(-1963./204120))))))));
  }


  template<>
  inline
  double
  BranchPointPolynomial<9>(const double p)
  {
    return
      -1 + p*(1 + p*(-1./3 + p*(11./72 + p*(-43./540 + p*(769./17280
      + p*(-221./8505 + p*(680863./43545600 + p*(-1963./204120
      + p*(226287557./37623398400.)))))))));
  }


  template<int order>
  inline double AsymptoticExpansion(const double a, const double b);


  template<>
  inline
  double
  AsymptoticExpansion<0>(const double a, const double b)
  {
    return a - b;
  }


  template<>
  inline
  double
  AsymptoticExpansion<1>(const double a, const double b)
  {
    return a - b + b / a;
  }


  template<>
  inline
  double
  AsymptoticExpansion<2>(const double a, const double b)
  {
    const double ia = 1 / a;
    return a - b + b / a * (1 + ia * 0.5*(-2 + b));
  }


  template<>
  inline
  double
  AsymptoticExpansion<3>(const double a, const double b)
  {
    const double ia = 1 / a;
    return a - b + b / a *
      (1 + ia *
        (0.5*(-2 + b) + ia *
           1/6.*(6 + b*(-9 + b*2))
        )
      );
  }


  template<>
  inline
  double
  AsymptoticExpansion<4>(const double a, const double b)
  {
    const double ia = 1 / a;
    return a - b + b / a *
      (1 + ia *
        (0.5*(-2 + b) + ia *
          (1/6.*(6 + b*(-9 + b*2)) + ia *
            1/12.*(-12 + b*(36 + b*(-22 + b*3)))
          )
        )
      );
  }


  template<>
  inline
  double
  AsymptoticExpansion<5>(const double a, const double b)
  {
    const double ia = 1 / a;
    return a - b + b / a *
      (1 + ia *
        (0.5*(-2 + b) + ia *
          (1/6.*(6 + b*(-9 + b*2)) + ia *
            (1/12.*(-12 + b*(36 + b*(-22 + b*3))) + ia *
              1/60.*(60 + b*(-300 + b*(350 + b*(-125 + b*12))))
            )
          )
        )
      );
  }


  template<int branch>
  class Branch {

  public:
    template<int order>
    static double BranchPointExpansion(const double x)
    { return BranchPointPolynomial<order>(eSign * sqrt(2*(M_E*x+1))); }

    // Asymptotic expansion
    // Corless et al. 1996, de Bruijn (1981)
    template<int order>
    static
    double
    AsymptoticExpansion(const double x)
    {
      const double logsx = log(eSign * x);
      const double logslogsx = log(eSign * logsx);
      return LambertWDetail::AsymptoticExpansion<order>(logsx, logslogsx);
    }

    template<int n>
    static inline double RationalApproximation(const double x);

    // Logarithmic recursion
    template<int order>
    static inline double LogRecursion(const double x)
    { return LogRecursionStep<order>(log(eSign * x)); }

    // generic approximation valid to at least 5 decimal places
    static inline double Approximation(const double x);

  private:
    // enum { eSign = 2*branch + 1 }; this doesn't work on gcc 3.3.3
    static const int eSign = 2*branch + 1;

    template<int n>
    static inline double LogRecursionStep(const double logsx)
    { return logsx - log(eSign * LogRecursionStep<n-1>(logsx)); }
  };


  // Rational approximations

  template<>
  template<>
  inline
  double
  Branch<0>::RationalApproximation<0>(const double x)
  {
    return x*(60 + x*(114 + 17*x)) / (60 + x*(174 + 101*x));
  }


  template<>
  template<>
  inline
  double
  Branch<0>::RationalApproximation<1>(const double x)
  {
    // branch 0, valid for [-0.31,0.3]
    return
      x * (1 + x *
        (5.931375839364438 + x *
          (11.392205505329132 + x *
            (7.338883399111118 + x*0.6534490169919599)
          )
        )
      ) /
      (1 + x *
        (6.931373689597704 + x *
          (16.82349461388016 + x *
            (16.43072324143226 + x*5.115235195211697)
          )
        )
      );
  }


  template<>
  template<>
  inline
  double
  Branch<0>::RationalApproximation<2>(const double x)
  {
    // branch 0, valid for [-0.31,0.5]
    return
      x * (1 + x *
        (4.790423028527326 + x *
          (6.695945075293267 + x * 2.4243096805908033)
        )
      ) /
      (1 + x *
        (5.790432723810737 + x *
          (10.986445930034288 + x *
            (7.391303898769326 + x * 1.1414723648617864)
          )
        )
      );
  }


  template<>
  template<>
  inline
  double
  Branch<0>::RationalApproximation<3>(const double x)
  {
    // branch 0, valid for [0.3,7]
    return
      x * (1 + x *
        (2.4450530707265568 + x *
          (1.3436642259582265 + x *
            (0.14844005539759195 + x * 0.0008047501729129999)
          )
        )
      ) /
      (1 + x *
        (3.4447089864860025 + x *
          (3.2924898573719523 + x *
            (0.9164600188031222 + x * 0.05306864044833221)
          )
        )
      );
  }


  template<>
  template<>
  inline
  double
  Branch<-1>::RationalApproximation<4>(const double x)
  {
    // branch -1, valid for [-0.3,-0.05]
    return
      (-7.814176723907436 + x *
        (253.88810188892484 + x * 657.9493176902304)
      ) /
      (1 + x *
        (-60.43958713690808 + x *
          (99.98567083107612 + x *
            (682.6073999909428 + x *
              (962.1784396969866 + x * 1477.9341280760887)
            )
          )
        )
      );
  }


  // stopping conditions

  template<>
  template<>
  inline
  double
  Branch<0>::LogRecursionStep<0>(const double logsx)
  {
    return logsx;
  }


  template<>
  template<>
  inline
  double
  Branch<-1>::LogRecursionStep<0>(const double logsx)
  {
    return logsx;
  }


  template<>
  inline
  double
  Branch<0>::Approximation(const double x)
  {
    if (x < -0.32358170806015724) {
      if (x < -kInvE)
        return numeric_limits<double>::quiet_NaN();
      else if (x < -kInvE+1e-5)
        return BranchPointExpansion<5>(x);
      else
        return BranchPointExpansion<9>(x);
    } else {
      if (x < 0.14546954290661823)
        return RationalApproximation<1>(x);
      else if (x < 8.706658967856612)
        return RationalApproximation<3>(x);
      else
        return AsymptoticExpansion<5>(x);
    }
  }


  template<>
  inline
  double
  Branch<-1>::Approximation(const double x)
  {
    if (x < -0.051012917658221676) {
      if (x < -kInvE+1e-5) {
        if (x < -kInvE)
          return numeric_limits<double>::quiet_NaN();
        else
          return BranchPointExpansion<5>(x);
      } else {
        if (x < -0.30298541769)
          return BranchPointExpansion<9>(x);
        else
          return RationalApproximation<4>(x);
      }
    } else {
      if (x < 0)
        return LogRecursion<9>(x);
      else if (x == 0)
      return -numeric_limits<double>::infinity();
      else
        return numeric_limits<double>::quiet_NaN();
    }
  }


  // iterations

  inline
  double
  HalleyStep(const double x, const double w)
  {
    const double ew = exp(w);
    const double wew = w * ew;
    const double wewx = wew - x;
    const double w1 = w + 1;
    return w - wewx / (ew * w1 - (w + 2) * wewx/(2*w1));
  }


  inline
  double
  FritschStep(const double x, const double w)
  {
    const double z = log(x/w) - w;
    const double w1 = w + 1;
    const double q = 2 * w1 * (w1 + (2/3.)*z);
    const double eps = z / w1 * (q - z) / (q - 2*z);
    return w * (1 + eps);
  }


  template<
    double IterationStep(const double x, const double w)
  >
  inline
  double
  Iterate(const double x, double w, const double eps = 1e-6)
  {
    for (int i = 0; i < 100; ++i) {
      const double ww = IterationStep(x, w);
      if (fabs(ww - w) <= eps)
        return ww;
      w = ww;
    }
    cerr << "convergence not reached." << endl;
    return w;
  }


  template<
    double IterationStep(const double x, const double w)
  >
  struct Iterator {

    static
    double
    Do(const int n, const double x, const double w)
    {
      for (int i = 0; i < n; ++i)
        w = IterationStep(x, w);
      return w;
    }

    template<int n>
    static
    double
    Do(const double x, const double w)
    {
      for (int i = 0; i < n; ++i)
        w = IterationStep(x, w);
      return w;
    }

    template<int n, class = void>
    struct Depth {
      static double Recurse(const double x, double w)
      { return Depth<n-1>::Recurse(x, IterationStep(x, w)); }
    };

    // stop condition
    template<class T>
    struct Depth<1, T> {
      static double Recurse(const double x, const double w)
      { return IterationStep(x, w); }
    };

    // identity
    template<class T>
    struct Depth<0, T> {
      static double Recurse(const double x, const double w)
      { return w; }
    };

  };


} // LambertWDetail


template<int branch>
double
LambertWApproximation(const double x)
{
  return LambertWDetail::Branch<branch>::Approximation(x);
}

// instantiations
template double LambertWApproximation<0>(const double x);
template double LambertWApproximation<-1>(const double x);


template<int branch>
double LambertW(const double x);

template<>
double
LambertW<0>(const double x)
{
  if (fabs(x) > 1e-6 && x > -LambertWDetail::kInvE + 1e-5)
    return
      LambertWDetail::
        Iterator<LambertWDetail::FritschStep>::
          Depth<1>::
            Recurse(x, LambertWApproximation<0>(x));
  else
    return LambertWApproximation<0>(x);
}

template<>
double
LambertW<-1>(const double x)
{
  if (x > -LambertWDetail::kInvE + 1e-5)
    return
      LambertWDetail::
        Iterator<LambertWDetail::FritschStep>::
          Depth<1>::
            Recurse(x, LambertWApproximation<-1>(x));
  else
    return LambertWApproximation<-1>(x);
}

// instantiations
template double LambertW<0>(const double x);
template double LambertW<-1>(const double x);
\end{verbatim}
}

\end{multicols}

\end{document}